# Characteristics of Spreadsheets Developed with the SSMI Methodology

*Paul Mireault*
*Founder, SSMI International*
*Honorary Professor, HEC Montréal*
*Paul.Mireault@SSMI.International*

**ABSTRACT**

*The SSMI methodology was developed using concepts from Computer Science, Software Engineering and Information Systems and has been taught to undergraduate and MBA students and in Executive training seminars. In this paper, we describe the major characteristics of the spreadsheets developed using the methodology and show how they contribute to reduce many error causing factors.*

## 1 INTRODUCTION

Developing a spreadsheet is a complex task, often performed by people with little or no training. These may be accountants, analysts or department directors who have a well-defined job to do, which is not being a *spreadsheet specialist*. The free-form nature of spreadsheets lead to all sorts of designs.

The SSMI methodology was developed to help developers structure their spreadsheet in a way that makes them easy to understand and to maintain.

## 2 REVIEW OF THE SSMI METHODOLOGY

The Structured Spreadsheet Modelling and Implementation (SSMI) methodology is described in (Mireault, 2016). It uses the following concepts that are commonly used in domains related to systems development.

1. **Conceptual model**. In Information Systems, the conceptual model is used to describe what a system does, or should do, without considering the technology that will be used to implement it. It uses a vocabulary familiar to the user.
2. **Names**. In Computer Science all computer languages use symbolic names to indicate what variables represent. The only restriction, in Excel, is that names cannot contain spaces and some other special characters. When Excel creates names from cell labels, it replaces spaces and special characters with the underscore "_" character.
3. **Modules**. In Computer Science, modules are self-contained portions of code that have a specific list of inputs and produce a specific output. Modules are easier to understand and to debug.
4. **3-tier architecture**. In Software Engineering, the 3-tier architecture consists of separating an implementation in elements that handle different operations. The usual tiers are the *Interface*, the *Application* and the *Services*. With this separation of major tasks, one can change database systems by modifying only the *Services* tier and leaving the other tiers untouched.

In the SSMI methodology, the conceptual model consists of the *Formula Diagram* which is a representation of the problem's variables and how they are related, and the *Formula List* which specifies the nature of these relationships as formulas. Figure 1, taken from (Mireault, 2016), illustrates a *Formula Diagram* and Table 1 its *Formula List*. In this example, the user wants to examine scenarios where he modifies the product's price and see the corresponding regional Profit and Total Profit. He uses a demand function to estimate the total Demand for a given Price. The dash-bordered rectangle indicates the portion of the model that depends on the Region dimension: each region has its own delivery cost and the Distribution parameter represents its portion of the overall Demand.

Rectangles represent values that will be entered by the user during the normal use of the spreadsheet and triangles represent constants that are not usually changed. Circles and ovals represent variables that are calculated with a formula, with ovals representing variables whose value are of interest to the user. Rectangles and ovals will be part of the *Interface* tier.

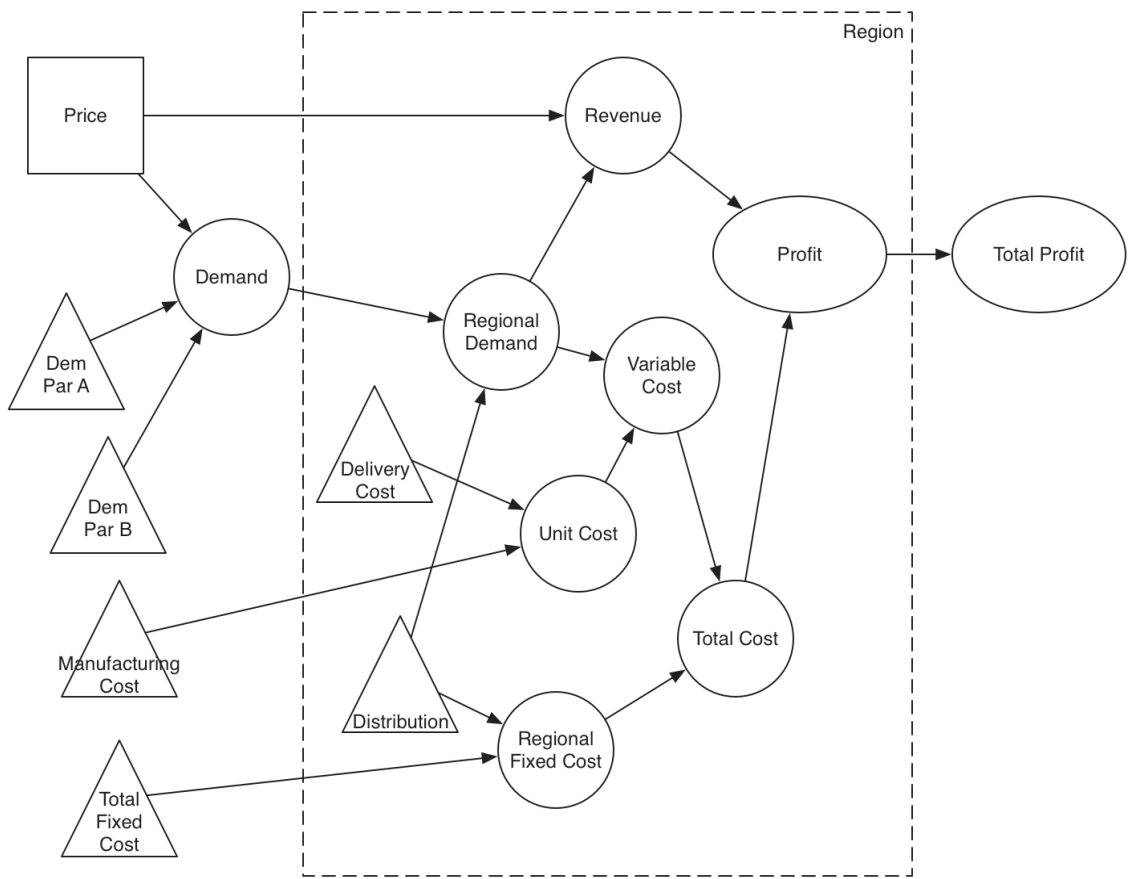

*Figure 1 - Formula Diagram*

| Variable | Type | Definition |
|---|---|---|
| Price | Input | (To be set by user) |
| Profit | Output, repeating | Revenue – Total Cost |
| DemParA | Parameter | 376,000 |
| DemParB | Parameter | 1.009 |
| Fixed Cost | Parameter | $2,500,000 |
| Manufacturing Cost | Parameter | $120 |
| Distribution | Parameter, repeating | 48%, 23%, 29% |
| Delivery Cost | Parameter, repeating | $50, $80, $60 |
| Total Demand | Calculated | DemParA * DemParB^–Price |
| Regional Demand | Calculated, repeating | Total Demand * Distribution |

| Variable | Type | Definition |
|---|---|---|
| Total Cost | Calculated, repeating | Regional Fixed Cost + Variable Cost |
| Regional Fixed Cost | Calculated, repeating | Fixed Cost * Distribution |
| Variable Cost | Calculated, repeating | Regional Demand * Unit Cost |
| Unit Cost | Calculated, repeating | Manufacturing Cost + Delivery Cost |
| Revenue | Calculated, repeating | Regional Demand * Price |
| Total Profit | Output | SUM(Profit) |

Table 1 - Formula List

Names are implemented directly using Excel's *Create Name from Selection* button (or menu item), as shown in Figure 2. Names created in the structured implementation refer to either single cells, entire row or entire columns.

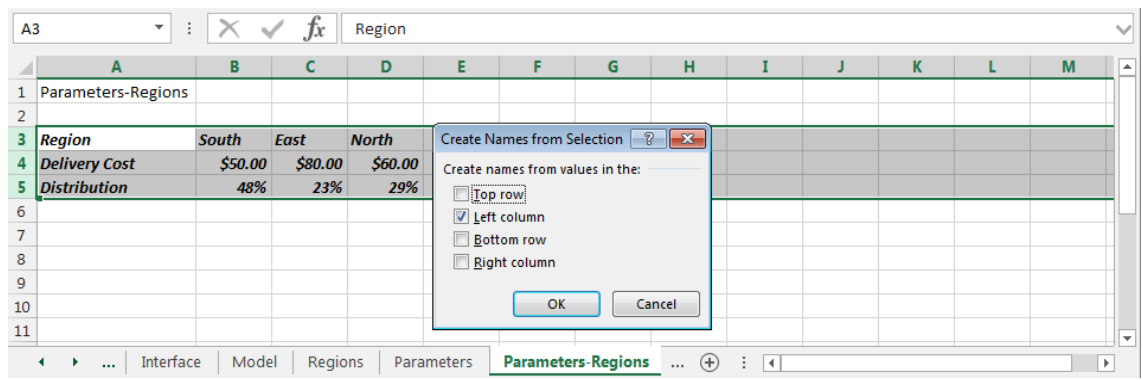

Figure 2 - Creating Names using the row labels

In the structured implementation, each calculated variable is defined at the bottom of its *definition block*, as illustrated in Figure 3.

|   | A | B | C | D |
|---|---|---|---|---|
| 5 | Total Demand | 13,062 | 13,062 | 13,062 |
| 6 | Distribution | 48% | 23% | 29% |
| 7 | **Regional Demand** | **6,270** | **3,004** | **3,788** |
| 8 | | | | |
| 9 | Regional Demand | 6,270 | 3,004 | 3,788 |
| 10 | Price | $375.00 | $375.00 | $375.00 |
| 11 | **Revenue** | **$2,351,110.34** | **$1,126,573.70** | **$1,420,462.50** |

Figure 3 - Definition blocks of variables Regional Demand and Revenue

The top part of a definition block consists only in *reference formulas* whose purpose is to make a local copy the values used in the definition formula, as shown in Figure 4.

|   | A | B | C | D |
|---|---|---|---|---|
| 5 | Total Demand | =Total_Demand | =Total_Demand | =Total_Demand |
| 6 | Distribution | =Distribution | =Distribution | =Distribution |
| 7 | **Regional Demand** | =B5*B6 | =C5*C6 | =D5*D6 |
| 8 | | | | |
| 9 | Regional Demand | =Regional_Demand | =Regional_Demand | =Regional_Demand |
| 10 | Price | =Price | =Price | =Price |
| 11 | **Revenue** | =B9*B10 | =C9*C10 | =D9*D10 |

Figure 4 - Formula view showing the structure of a definition block

The SSMI methodology can be taught in 8 to 10 hours (Mireault, 2016).

In the following section, we will describe the major characteristics of the spreadsheets developed by following the SSMI methodology.

## 3 SSMI CHARACTERISTICS

### 3.1 Use of worksheets

The SSMI methodology uses worksheets to implement the 3-tier architecture. Worksheets are dedicated to a single tier:

- The Application tier: the ***Model*** worksheets contain only the definition blocks of the model's variables. There are never any input cells in the model worksheets. Also, the model worksheets are not designed for the worksheet user: the layout elements are present only to help the worksheet developer of the worksheet auditor. The definition of a variable is indicated by a bold-italic font. There can be more than one Model worksheet to suit the developer's needs.
- The Services tier: the ***Parameters*** worksheets are the sheets that collect all the inputs. The inputs can come from the Interface sheet or from sheets that import raw data from external sources.
- The Interface tier: the ***Interface*** sheets are the sheets that are actually used by the spreadsheet's users. This is where they enter specific values and see the results for the scenarios that interest them. The input values are referenced in the Parameters worksheet and the output values are referenced from the Models worksheets.

Using dedicated worksheets is also recommended by many researchers, such as (Read & Batson, 1999).

### 3.2 Use of Names

Using names indiscriminately has been identified as a source of error. (McKeever & McDaid, 2010) devised an experiment whose results suggest that names has a negative effect on debugging performance. (Kruck & Sheetz, 2001) suggest that naming cells contribute to making formulas easier to understand Nonetheless, we consider that the way we use names is very different than the unstructured way that was used in those studies.

The structured implementation of the SSMI methodology uses names only in the reference formulas that are part of the upper portion of the definition blocks. The definition formula itself does not use names. This way, we take advantage of Excel's color coding to help us validate the definition formula: as shown in Figure 5, each element of the formula in cell B6 has a colour code and each element of the top portion of the definition block also has a colour code. If one element did not have a colour code, then we would investigate further.

*Figure 5 – Excel's color coding in a definition formula*

### 3.3 Far and local references

Far references in formulas, called *coupling by* (Kruck & Sheetz, 2001), has been identified as a source of errors. (Raffensperger, 2003) says that a formula that references a cell that is not immediately visible and understood is harder to understand,

A definition formula uses local references to values that are situated in the top part of its definition block. This way, far references are not specified by point-and-click; they are specified by the name of the variable. An error could occur when the developer types the wrong name, but since we display the variable's values to the right of its label the developer should notice the error. Figure 4 shows that it is easy to verify that the proper reference is made in the top portion of the definition blocks.

### 3.4 Transitive references

A *transitive reference* is a reference to a location where a variable is *used*, not where it is *defined*.

The left side of Figure 6 illustrates a situation where the developer refers to cell B10 in cell B14 instead of referring to the definition of *Number of Items Delivered* (cell B3). Later, he modifies the spreadsheet to take into account the fact that all the items delivered incur a delivery cost but only the items effectively sold contribute to the total sales. He does the correct modification in rows 10 to 12, but he does not realize that the transitive reference in cell B14 now produces an error.

While a transitive reference does not produce an error when the spreadsheet is initially built, it can produce a logical error when the spreadsheet is modified later. Spreadsheet developers are tempted to use transitive references because it is faster to point to a cell that is close (usually just above) than to navigate to the location where the variable is actually defined (which can be far).

Transitive references have been identified as a cause of errors by many authors. The errors can appear during the spreadsheet maintenance when new variables are created to take new nuances into account.

*Figure 6 - Example of a transitive reference*

Using the SSMI methodology, the developer simply cannot create a transitive reference since the references in the top portion of a calculation block use names that always refer to the definition of the variable. This illustrated in Figure 7: the left side shows the formula view of the left side of Figure 6 developed without the SSMI methodology and the right side shows the formula view of the spreadsheet developed with the SSMI methodology.

*Figure 7 - The SSMI methodology cannot produce a transitive reference*

Figure 8 illustrates the formula view of the right side of Figure 6.

|   | A | B |
|---|---|---|
| 1 | Unit Price | 12 |
| 2 | Unit Delivery Cost | 8 |
| 3 | Number of Items Delivered | 1000 |
| 4 | Number of Items Returned | 50 |
| 5 |  |  |
| 6 | Number of Items Delivered | =Number_of_Items_Delivered |
| 7 | Number of Items Returned | =Number_of_Items_Returned |
| 8 | Number of Items Sold | =B6-B7 |
| 9 |  |  |
| 10 | Number of Items Sold | =Number_of_Items_Sold |
| 11 | Unit Price | =Unit_Price |
| 12 | Total Sales | =B10*B11 |
| 13 |  |  |
| 14 | Number of Items Delivered | =Number_of_Items_Delivered |
| 15 | Unit Delivery Cost | =Unit_Delivery_Cost |
| 16 | Total Delivery Cost | =B14*B15 |

*Figure 8 - Formula view of the right side of Figure 6*

### 3.5  Formula Complexity

The complexity of the formula in a cell is often cited as a source of errors. According to (Hermans, et al., 2012) simpler formulas have a low complexity score.

The SSMI methodology's most important rule is to never mix operators or functions in a formula. The developer is encouraged to create variables as needed. The formula

$$\text{Total Cost} = \text{Fixed Cost} + \text{Quantity} * \text{Unit Cost}$$

uses two different operators: the addition and the multiplication. It is thus replaced by the two following formulas:

$$\text{Variable Cost} = \text{Quantity} * \text{Unit Cost}$$

$$\text{Total Cost} = \text{Fixed Cost} + \text{Variable Cost}$$

When implemented, this rule has the advantage of creating blocks that are easy to validate by eye, as shown in Figure 9.

| 5 | Fixed Cost | $12,638.00 |
| 6 | Unit Cost | $13.28 |
| 7 | Quantity | 5,287 |
| 8 | Total Cost | $82,849.36 |
| 9 | | |
| 10 | | |
| 11 | | |
| 12 | | |

| 5 | Unit Cost | $13.28 |
| 6 | Quantity | 5,287 |
| 7 | Variable Cost | $70,211.36 |
| 8 | | |
| 9 | Fixed Cost | $12,638.00 |
| 10 | Variable Cost | $70,211.36 |
| 11 | Total Cost | $82,849.36 |
| 12 | | |

*Figure 9 - Formula complexity example*

### 3.6 Formula copying

Copying formulas has been cited by many authors as a source of errors. The errors can be due to improper relative or absolute references or to partial copying.

Relative and absolute references are an artefact introduced by spreadsheet program companies, like Microsoft, Lotus and Apple, to understand the spreadsheet developer's intentions when he copies a formula.

In an SSMI spreadsheet, there is never any need to use absolute or mixed references in a formula. As mentioned above, the definition formula uses only the cells that are immediately above it, in the top portion of the block, and uses the standard relative references. The cells in the top part of the block always use names that are interpreted correctly. A name that refers to a single cell is interpreted as an absolute reference, and a name that refers to a row or a column is interpreted as a mixed reference. This behaviour is illustrated in Figure 10, showing the normal and the formula view of the same worksheet with the *Trace Precedents* arrows.

|   | A | B | C | D | E | F | G | H | I |
|---|---|---|---|---|---|---|---|---|---|
| 1 | Price | $325 | | | | | | | |
| 2 | | | | | | | | | |
| 3 | Region | South | East | North | | | | | |
| 4 | Regional Demand | 6,269 | 3,004 | 3,787 | | | | | |
| 5 | | | | | | | | | |
| 6 | Regional Demand | 6,269 | 3,004 | 3,787 | | | | | |
| 7 | Price | $325 | $325 | $325 | | | | | |
| 8 | Revenue | $2,037,425 | $976,300 | $1,230,775 | | | | | |
| 9 | | | | | | | | | |

|   | A | B | C | D |
|---|---|---|---|---|
| 1 | Price | 325 | | |
| 2 | | | | |
| 3 | Region | South | East | North |
| 4 | Regional Demand | 6269 | 3004 | 3787 |
| 5 | | | | |
| 6 | Regional Demand | =Regional_Demand | =Regional_Demand | =Regional_Demand |
| 7 | Price | =Price | =Price | =Price |
| 8 | Revenue | =B6*B7 | =C6*C7 | =D6*D7 |

*Figure 10 - Behaviour of Names. Normal view (top) and Formula view (bottom)*

A partial copy happens when the developer does not copy the cell's formula all the way to the end of the row or column. The result is that some cells in the row, or column, have one formula and the others have another.

In a repeating model worksheet, the model is implemented in a single column of its worksheet. Even if a few formulas have been modified, it is easier to copy the whole column instead of copying the modified formulas one at the time. Reducing the number of times the developer needs to copy formulas should reduce the risk of making copy errors.

### 3.7 Auditing

When we audit a spreadsheet we are in fact verifying two things: the model and its implementation.

**Understanding the model**

The first step consists of examining each variable of the Formula Diagram and determining if its formula is correct. During this step, we also have to determine if we have forgotten or ignored other variables.

**Verifying the spreadsheet**

Once we are satisfied that the model is correct, we now need to verify if its implementation is also correct. Verifying an SSMI spreadsheet is pretty straightforward. There are two cases to consider: the non-repeating model and the repeating model.

In the case of the non-repeating model we need only display the formula view in Excel and examine each definition block to see if it conforms to the Formula Diagram and the Formula List.

In the case of the repeating model we do the same examination on the leftmost column of the repeating model worksheet. Then, to make sure that no formula has been altered in the other columns or that there hasn't been a partial copy, we can proceed as follows.

1. Work on a copy of the file you are auditing.
2. In the repeating model worksheet, copy everything: Ctrl+A, Ctrl+C.
3. Create an empty worksheet and paste the values only. Select everything and create a conditional format highlighting the cells that have a different value than the corresponding cell in the repeating model worksheet.
4. Return to the repeating model worksheet and copy the first model column to all the others on the right.
5. Examine the worksheet with the pasted values to see if any cell is highlighted.

## 4  CONCLUSION

(Panko, 2015) says that "Given that [our] experience of errors in unreliable, intuitions about how to reduce errors should not be taken seriously unless they are rigorously tested." The SSMI methodology has not been tested yet. But neither have the other standards proposed by different organizations: FAST (Fast Standard Organization, 2015), SMART (Corality, 2015) and SSRB (Spreadsheet Standards Review Board, 2012). According to (Grossman & Özlük, 2010), these standards "do not attempt to address "writing spreadsheets" in general" but are specialized for financial modelling.

The SSMI methodology has been developed to be a general spreadsheet development methodology that can be used in any domain, like accounting, actuary, economics, engineering, human resources, insurance, logistics, management, marketing and, of course, finance.

Further research, like (McKeever & McDaid, 2011), should be done to evaluate its performance with regards to errors and ease of use,